\documentstyle[aps,prb,epsfig,float,twocolumn]{revtex}
\begin{document}
\twocolumn[\hsize\textwidth\columnwidth\hsize\csname
@twocolumnfalse\endcsname

\title{Correlation energy, quantum phase transition, and bias potential effects
in quantum Hall bilayers at $\nu=1$}

\author{John Schliemann}

\address{Department of Physics and Astronomy, University of Basel, 
CH-4056 Basel, Switzerland}

\date{\today}

\maketitle

\begin{abstract}
We study the correlation energy, the effective anisotropy parameter,
and quantum fluctuations of the pseudospin magnetization in
bilayer quantum Hall systems at total filling factor $\nu=1$
by means of exact diagonalizations of the Hamiltonian 
in the spherical geometry. We compare exact diagonalization 
results for the ground state energy with finite-size Hartree-Fock values.
In the ordered ground state phase at small layer separations the Hartree-Fock
data compare reasonably with the exact results. Above the critical layer
separation, however, the Hartree-Fock findings still predict an increase
in the ground state energy, while the exact ground state energy is in this
regime independent of the layer separation indicating the decoupling of layers
and the loss of spontaneous phase coherence between them. 
We also find accurate values for the
pseudospin anisotropy constant whose dependence of the layer separation
provides another very clear indication for the strong interlayer
correlations in the ordered phase and shows an inflection point at
the phase boundary.

Finally we discuss the possibility of interlayer correlations in biased
systems even above the phase boundary for the balanced case. Certain features
of our data for the pseudospin anisotropy constant as well as for
quantum fluctuations of the pseudospin magnetization 
are not inconsistent with the occurrence of this effect.
However, it appears to be rather weak at least in the limit of 
vanishing tunneling amplitude.
\end{abstract}
\vskip2pc]


\section{Introduction}

Quantum Hall ferromagnets are a rich and fascinating field of solid
state physics \cite{DasSarma97,Girvin00,Jungwirth01}. 
They can be realized in terms of the
spins of electrons confined to layers in a strong perpendicular magnetic field,
or in terms of a pseudospin given by some additional discrete degree
of freedom such as the layer spin in bilayer systems
\cite{Fertig89,MacDonald90,Cote92,He93,Yang94,Moon95,Yang96,Murphy94}.
Bilayer quantum Hall systems at total filling factor $\nu=1$ have attracted 
particular interest recently due to spectacular results by Spielman 
{\it et al.} \cite{Spielman00,reviews} 
who studied tunneling transport across 
the layers in samples with very small single-particle tunneling gap.
These experiments have stimulated a large number of theoretical
efforts toward their explanation, and also more general studies of
such bilayer quantum Hall systems
\cite{Balents01,Stern01,Fogler01,Schliemann01,Demler01,Yang01,Hanna97,Joglekar01a,Joglekar01b,Joglekar02,Hanna01,EYang01,Stern02,Nakajima01,Veillette01,Burkov02,Nomura02,Yu02,Abolfath02a,Abolfath02b}.

The main finding of Refs. \cite{Spielman00} is a pronounced peak in the
differential tunneling conductance which evolves if the 
layer separation $d$ in units of the magnetic length $\ell$ is decreased
below a critical value. This critical ratio $d/\ell$ agrees closely with
the boundary between a ground state phase supporting quantized Hall transport
and a disordered phase as established in earlier experiments by
Murphy {\it et al.} \cite{Murphy94}
using double well samples of similar geometry. Therefore these two
observations can be assumed to be manifestations of one and the same
quantum phase transition. Moreover, recent exact-diagonalization studies
\cite{Schliemann01}
on bilayers at $\nu=1$ have revealed a quantum phase transition, very likely
to be of first order, between a phase with strong interlayer correlations
to a phase with weak interlayer correlations. The position of this transition
agrees quantitatively with the critical value found by Spielman {\it et al.}

In the ordered phase at small $d/\ell$ the strong interlayer
correlations are dominated by the {\em spontaneous interlayer phase
coherence} between the layers. This key word describes the fact that 
in the ground state of such a system electrons predominantly occupy
single-particle states in the lowest Landau level which
are symmetric linear combination of states in both layers. This 
type of single-particle states is preferred if a finite tunneling
amplitude is present. However, by a large body of
experimental and theoretical work
\cite{Fertig89,MacDonald90,Cote92,He93,Yang94,Moon95,Yang96,Murphy94,Spielman00,reviews,Balents01,Stern01,Fogler01,Schliemann01,Demler01,Yang01,Hanna97,Joglekar01a,Joglekar01b,Joglekar02,Hanna01,EYang01,Stern02,Nakajima01,Veillette01,Burkov02,Nomura02,Yu02,Abolfath02a,Abolfath02b}, this phenomenon is assumed to be a spontaneous symmetry
breaking, i.e. it remains even in the limit of vanishing tunneling amplitude.
The latter effect is clearly a many-body phenomenon.

In the present work we report on further exact-diagonalization results in
quantum Hall bilayers at total filling factor $\nu=1$. Our studies include
the effective pseudospin anisotropy parameter, quantum fluctuations
of the pseudospin magnetization, and the ground state energy.
Especially the last quantity shows very clearly the occurrence of the
quantum phase transition and the decoupling of the layers above the
critical $d/\ell$ where the spontaneous interlayer phase coherence is lost.
Moreover, we study the effects of a bias potential applied
to the layers. 
In particular we address the question of possible interlayer correlations
in biased systems even above the phase boundary of the balanced case, an
effect which was predicted recently by Hanna \cite{Hanna97} and by 
Joglekar and MacDonald \cite{Joglekar02} based on
time-dependent Hartree-Fock calculations.
Some features of our data are not inconsistent with this prediction.
However, this effect appears to be rather weak  
at least in the limit of vanishing tunneling
amplitude and not too large biasing,  
consistent with the predictions of Refs.~\cite{Hanna97,Joglekar02}.

Our numerics are performed within the spherical
geometry \cite{Haldane83}. This geometry enables to obtain closed
expressions for the Hartree-Fock ground state energy even in finite
systems. This quantity can be compared with exact-diagonalization results
to infer the correlation energy. Moreover, since the sphere is free of
boundaries, this geometry allows to take into account a neutralizing
background in finite systems without any ambiguity.

This paper is organized as follows. In section \ref{HF} we describe
our finite-size Hartree-Fock calculations of the ground state energy
in the spherical geometry. In section \ref{results} we present our
exact-diagonalization results, compare them with Hartree-Fock
theory, and perform a detailed analysis of bias potential effects.
We close with conclusions in section \ref{conclusions}.


\section{Finite-size Hartree-Fock theory in the spherical geometry}
\label{HF}

In this section we present details of our finite-size Hartree-Fock
calculations in the spherical geometry \cite{Haldane83}. Similar results
for the case of bilayers at filling factor $\nu=2$ were already briefly
discussed in Ref.~\cite{Schliemann00}. The notation follows the discussion
of the $\nu=2$ system in Ref.~\cite{MacDonald99}. The technical advantage
of the spherical geometry used here lies in the fact that it allows to obtain
closed results for electron pair distribution function even in finite
systems. 

We consider a gas of Coulomb-interacting electrons in a
quantum Hall bilayer system at total filling factor $\nu=1$. We assume 
a vanishing amplitude for electron tunneling between the layers,
consistent with the experimental situation in Ref. \cite{Spielman00}.
The layer degree of freedom is described in the usual pseudospin language
\cite{DasSarma97} where the pseudospin operator of each electron is given by
$\vec\tau/2$ with $\vec\tau$ being the vector of Pauli matrices. 
The $z$-component 
$\tau^{z}/2$ measures the difference in occupation between the two layers,
while $\tau^{x}/2$ describes tunneling between them. The total
pseudospin of all electrons is denoted by $\vec T$.

Differently from the pseudospin,
the true electron spins are assumed to be fully aligned
along the magnetic field perpendicular to the layers; therefore an inessential
Zeeman term in the Hamiltonian is, along with the constant cyclotron
energy, neglected. In Refs. \cite{Schliemann01} a finite width of the
quantum wells forming the bilayer system was taken into account in order to
make quantitative contact to the experiments of Refs. \cite{Spielman00}.
However, a finite well width mainly changes the position of the
phase transition but does not alter any qualitative feature. In the present
work we therefore concentrate for simplicity on the case of zero
well width. For this case the critical layer separation in the limit of
vanishing tunneling amplitude was found by exact-diagonalization calculations 
\cite{Schliemann01} to be $d=1.3\ell$. This value holds in the thermodynamic
limit, but is remarkably rapidly approached in finite-size systems
\cite{Schliemann01}. For instance, the phase boundary in a system of
just 12 electrons deviates from the infinite-volume value
by just a few percent.

In the gauge commonly used in the spherical geometry \cite{Haldane83} the
single-particle wave functions in the lowest Landau level have the
form
\begin{eqnarray}
\langle\vec r|m\rangle
& = & \left(\frac{N_{\phi}+1}{2\pi\ell^{2}N_{\phi}}
{{N_{\phi}}\choose{\frac{N_{\phi}}{2}+m}}\right)^{\frac{1}{2}}
\left(\cos\left(\frac{\vartheta}{2}\right)
e^{i\frac{\varphi}{2}}\right)^{\frac{N_{\phi}}{2}+m}\nonumber\\
 & & \quad\quad\cdot\left(\sin\left(\frac{\vartheta}{2}\right)
e^{-i\frac{\varphi}{2}}\right)^{\frac{N_{\phi}}{2}-m}
\end{eqnarray}
where $\vartheta$, $\varphi$ are the usual angular coordinates 
of the location $\vec r$ on the sphere with radius 
$|\vec r|=\ell\sqrt{N_{\phi}/2}$.
$m\in\{-\frac{N_{\phi}}{2},-\frac{N_{\phi}}{2}+1,\dots,
\frac{N_{\phi}}{2}\}$ labels the different angular momentum states, 
and $N_{\phi}$
is the number of flux quanta penetrating the sphere. The Hartree-Fock ansatz
for a spatially homogeneous state of $N=N_{\phi}+1$ electrons is 
\begin{equation}
|\Psi\rangle=\prod_{m=-\frac{N_{\phi}}{2}}^{\frac{N_{\phi}}{2}}
\left(\sum_{\sigma\in\{T,B\}}z_{\sigma}c^{+}_{m\sigma}\right)
|0\rangle
\end{equation}
where $|0\rangle$ is the fermionic vacuum. $c^{+}_{m\sigma}$,
$\sigma\in\{T,B\}$, creates an electron in the top/bottom layer in angular 
momentum state $m$, and $z_{\sigma}$ are the components of a normalized
two-spinor describing the layer degree of freedom. From this state we obtain
the pair distribution functions
\begin{eqnarray}
 g(\vec r_{1}-\vec r_{2})=
\langle\Psi|\sum_{i\neq j}
\left[\delta
(\vec r_{1}-\hat{\vec r_{i}})\delta(\vec r_{2}-\hat{\vec r_{j}})\right]
|\Psi\rangle & &\nonumber\\
 = \left(\frac{N_{\phi}+1}{2\pi\ell^{2}N_{\phi}}\right)^{2}
\left(1-\left(1-\frac{|\vec r_{1}-\vec r_{2}|^{2}}{2\ell^{2}N_{\phi}}
\right)^{N_{\phi}}\right) & &
\label{pairdis1}
\end{eqnarray}
\begin{eqnarray}
h(\vec r_{1}-\vec r_{2}) & = & 
\langle\Psi|\sum_{i\neq j}
\left[\delta(\vec r_{1}-\hat{\vec r_{i}})\tau_{i}^{z}
\delta(\vec r_{2}-\hat{\vec r_{j}})\tau_{j}^{z}\right]
|\Psi\rangle\nonumber\\
& = & \left(\frac{N_{\phi}+1}{2\pi\ell^{2}N_{\phi}}\right)^{2}
\left(\langle z|\tau^{z}|z\rangle\right)^{2}\nonumber\\
& & \cdot\left(1-\left(1-\frac{|\vec r_{1}-\vec r_{2}|^{2}}{2\ell^{2}N_{\phi}}
\right)^{N_{\phi}}\right) 
\label{pairdis2}
\end{eqnarray}
Here the indices $i,j$ refer to electrons and the Pauli matrices
$\tau^{z}$ act on the layer spins. The expression $|\vec r_{1}-\vec r_{2}|$
is the chord distance on the sphere. Note that in the limit of large
numbers of flux quanta $N_{\phi}$ one obtains from (\ref{pairdis1}) the
well-known expression for the infinite system in planar geometry,
\begin{equation}
\lim_{N_{\phi}\to\infty}g(r)=\left(\frac{1}{2\pi\ell^{2}}\right)^{2}
\left(1-e^{-\frac{r^{2}}{2\ell^{2}}}\right)
\end{equation}
To calculate the energy of the Coulomb interaction it is convenient
to consider the linear combination $V_{\pm}=(V_{S}\pm V_{D})/2$ of
the interactions $V_{S}$ and $V_{D}$ between electrons in the same layer
and different layers, respectively \cite{MacDonald99}. Using the above
pair distribution functions one obtains for the energy per particle
\begin{eqnarray}
\varepsilon^{HF} & = & \varepsilon^{HF}_{el}-\frac{1}{2}B\nonumber\\
& = & \frac{1}{2}\left(-F_{+}+\left(\langle z|\tau^{z}|z\rangle\right)^{2}
\left(H-F_{-}\right)\right)
\end{eqnarray}
Here $\varepsilon^{HF}_{el}$ is the Hartree-Fock energy of the interaction 
between electrons. The quantity
\begin{eqnarray}
B & = & \frac{e^{2}}{\epsilon\ell}\frac{N_{\phi}+1}{2\sqrt{N_{\phi}/2}}
\nonumber\\
 & & \cdot\left(1-\frac{1}{2\sqrt{N_{\phi}/2}}\frac{d}{\ell}
+\left(1+\frac{1}{N_{\phi}}\frac{d^{2}}{2\ell^{2}}\right)^{\frac{1}{2}}\right)
\label{background}
\end{eqnarray}
arises from the direct (Hartree) contribution of $V_{+}$ and cancels against
a neutralizing homogeneous background of half the total
electron charge which is present in each layer and ensures charge neutrality.
In this work we have always subtracted this term from the ground state
energies considered here. The quantity
\begin{eqnarray}
H & = & \frac{e^{2}}{\epsilon\ell}\frac{N_{\phi}+1}{2\sqrt{N_{\phi}/2}}
\nonumber\\
 & & \cdot\left(1+\frac{1}{2\sqrt{N_{\phi}/2}}\frac{d}{\ell}
-\left(1+\frac{1}{N_{\phi}}\frac{d^{2}}{2\ell^{2}}\right)^{\frac{1}{2}}\right)
\end{eqnarray}
stems from the direct term of $V_{-}$, and 
\begin{equation}
F_{\pm}=\frac{e^2}{\epsilon\ell}\frac{N_{\phi}+1}{2\sqrt{2N_{\phi}}}
\left(I(1)\pm
\left(\frac{1}{\alpha}\right)^{N_{\phi}+\frac{1}{2}}I(\alpha)\right)
\label{Fpm}
\end{equation}
represent the exchange (Fock) contributions from $V_{\pm}$ with
\begin{equation}
I(\alpha)=\int_{0}^{\alpha}\,dx\frac{x^{N_{\phi}}}{\sqrt{1-x}}
\quad,\quad
\alpha=\frac{1}{1+\frac{1}{N_{\phi}}\frac{d^2}{2\ell^{2}}}
\end{equation}
In the above equations, $e^{2}/(\epsilon\ell)$ is the Coulomb energy
scale with $(-e)$ being the electron charge and $\epsilon$ the dielectric
constant of the semiconductor material.
Note that all the above contributions to $\varepsilon^{HF}$ depend on the
layer separation $d/\ell$ as well as on the number of flux quanta $N_{\phi}$,
i.e. on the system size.

In the Hartree-Fock ground state of an unbiased system
all spins lie in the $xy$-plane of the
pseudospin space, i.e. $\langle z|\tau^{z}|z\rangle=0$, and we end up with
\begin{equation}
\varepsilon^{HF}_{0}=-\frac{1}{2}F_{+}
\end{equation}


\section{Results}
\label{results}

In this section we report on our results from exact numerical diagonalizations
of the many-body Coulomb Hamiltonian in the spherical geometry 
\cite{Haldane83}. In such a system the ground state has vanishing total angular
momentum \cite{Haldane83} and, for unbiased bilayers, 
the smallest possible value of the $z$-component of the total 
pseudospin $\vec T$, i.e. $T^{z}=0$ 
for an even number of electrons and $|T^{z}|=1/2$ otherwise. 

\subsection{Ground state and correlation energy in the unbiased system}
\label{gsunbiased}

\begin{figure}
\centerline{\includegraphics[width=8cm]{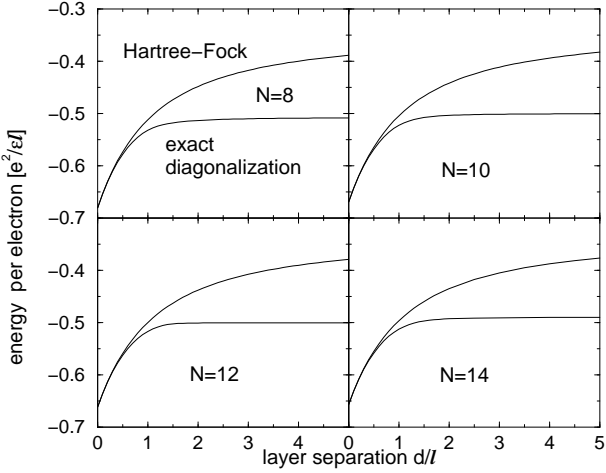}}
\caption{The ground state energy as a function of the layer separation
in units of the magnetic length 
for different numbers $N$ of electrons. The exact diagonalization 
data is compared with finite-size Hartree-Fock results. In both cases the
contribution from the neutralizing background has been subtracted.
In the ordered phase below the critical value of $d/\ell$ the results
agree reasonably and coincide for vanishing layer separation.
Above the critical layer separation the exact ground state energy is 
independent of $d/\ell$ corresponding to uncoupled $\nu=1/2$
monolayers, while Hartree-Fock theory still predicts an artificial increase
in ground state energy.
\label{fig1}}
\end{figure}

Figure \ref{fig1} shows the exact and the Hartree-Fock ground state energy
(both in units of the Coulomb energy scale $e^{2}/(\epsilon\ell)$) as a
function of $d/\ell$ for several numbers of electrons $N$. In both cases
the contribution from the neutralizing background (\ref{background})
is subtracted.
At zero layer separation we recover the case of a quantum Hall monolayer 
with the layer spin playing the role of the electron spin. 
Here the ground state is the well-known
spin-polarized $\nu=1$ monolayer ground state described exactly by
Hartree-Fock theory. In the spherical geometry the finite-size 
ground state energy per particle is given by
\begin{equation}
\varepsilon^{HF}=-\frac{e^{2}}{\epsilon\ell}
\frac{2^{2N_{\phi}}}{\sqrt{N_{\phi}/2}
{{2N_{\phi}+2}\choose{N_{\phi}+1}}}
\,{{{N_{\phi}\to\infty}\atop{\longrightarrow}}\atop{}}\,
-\frac{e^{2}}{\varepsilon\ell}\sqrt{\frac{\pi}{8}}
\end{equation}
with $N_{\phi}=N-1$ being the number of flux quanta.

At finite layer separation the Hartree-Fock ground state becomes unexact
but provides still a reasonable approximation to the exact ground state 
energy if $d/\ell$ is smaller than the critical value of $d/\ell=1.3$.
In other words, the correlation energy given by the difference between
the exact ground state energy and the Hartree-Fock value is small.
For larger layer separations $d/\ell\gtrsim1.3$ Hartree-Fock theory still
predicts an increase of the ground state energy with increasing layer
separation while the exact ground state energy becomes independent
of $d/\ell$. The latter result is again a particularly clear signature
of the decoupling of the two layers and the loss of spontaneous phase
coherence between them above the critical layer separation.
The discrepancy between the exact ground state energy and the Hartree-Fock
result in the disordered phase, i.e. the large correlation energy,
shows that this quantum phase transition is a correlation phenomenon that
cannot be described within simple Hartree-Fock theory.
In the Hartree-Fock {\em ansatz} used here all electrons are
in the same pseudospin state implementing phase coherence between the layers.
This coherence is lost above the critical $d/\ell$, and the system behaves,
at least in terms of its ground state energy, like two decoupled
monolayers with filling factor $\nu=1/2$.
Therefore, the failure of Hartree-Fock theory might appear as a consequence
of the artificial phase coherence. However, as it is well-known,  
the Hartree-Fock approach is generally inadequate to describe
quantum Hall monolayers at $\nu=1/2$, which have a very peculiar and
highly correlated ground state.

\subsection{The pseudospin anisotropy parameter and bias potential
effects}

The difference in the Coulomb interaction for electrons in the same 
layer and in different layers provides a strong mechanism balancing
the charges between the layers. In the pseudospin language this can be
expressed approximately by an effective easy-plane anisotropy 
contribution \cite{Moon95} to the energy per particle,
\begin{equation}
\varepsilon_{a}=8\pi\ell^{2}\beta\frac{\langle T^{z}\rangle^{2}}{N^{2}}
\label{anisoerg}
\end{equation}
introducing an anisotropy parameter $\beta$, and $\langle T^{z}\rangle$
denotes the expectation value of the $z$-component of the total pseudospin
\cite{note1}.
For vanishing tunneling between the layers as considered here 
this operator represents a
good quantum number, and eigenstates can be labeled their value of $T^{z}$.
In this case the above energy contribution can be viewed just as a charging
energy of a capacitor formed by the two isolated layers. In the absence of
quantum correlations, and for a large system, the anisotropy parameter
takes the value
\begin{equation}
8\pi\ell^{2}\beta_{cl}=\frac{e^{2}}{\epsilon\ell}\frac{d}{\ell}
\label{infclassbeta}
\end{equation}
corresponding to the classical total 
charging energy of $E_{c}=N\varepsilon_{a}=Q^{2}/(2C)$
with $Q=-eT^{z}$ being the charge of the capacitor, 
$C=\epsilon A/(4\pi d)$ its capacity, and $A=2\pi\ell^{2}N$ its area.
In the presence of quantum correlations the effective anisotropy
parameter will deviate from this value for two different reasons:\\
(i) {\em Interlayer correlations} can modify the value of $\beta$, and\\
(ii) even in the absence of correlations {\em between} the layers,
{\em intralayer correlations} 
can have an impact on $\beta$ if the ground states
of the two mutually uncorrelated layers change non-trivially if electrons are 
transferred from one layer to the other, i.e. if $T^{z}$ is changed. The latter
effect is independent of the layer separation. Therefore, in the absence
of interlayer correlations and for a given value of $T^{z}$, the 
contribution to the ground state energy which depends on the layer separation
is just given by a simple classical 
electrostatic expression \cite{note2} which can be derived
similarly as Eq.~\ref{background},
\begin{equation}
\varepsilon_{a}^{cl}=
8\pi\ell^{2}\beta_{cl}\frac{\langle T^{z}\rangle^{2}}{N^{2}}
\label{classanisoerg}
\end{equation}
with
\begin{eqnarray}
8\pi\ell^{2}\beta_{cl} & = & 
\frac{e^{2}}{\epsilon\ell}\frac{N_{\phi}+1}{\sqrt{N_{\phi}/2}}
\nonumber\\
 & & \cdot\left(1+\frac{1}{2\sqrt{N_{\phi}/2}}\frac{d}{\ell}
-\left(1+\frac{1}{N_{\phi}}\frac{d^{2}}{2\ell^{2}}\right)^{\frac{1}{2}}\right)
\label{classbeta}
\end{eqnarray}
which converges to the expression (\ref{infclassbeta}) for
$N=N_{\phi}+1\to\infty$. Thus, if no interlayer correlations are present,
the contribution to the effective anisotropy parameter with a nontrivial
dependence on the layer separation is given by the above classical
expression, with a possible additional contribution independent of the
layer separation which arises from intralayer quantum effects.

\begin{figure}
\centerline{\includegraphics[width=8cm]{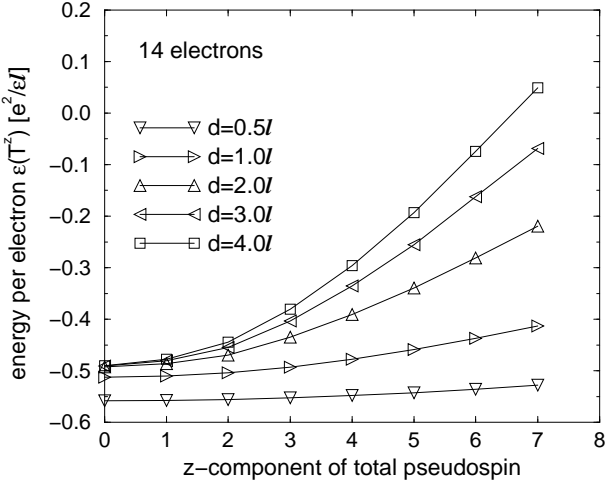}}
\caption{The energy of the lowest state having a given quantum number $T^{z}$
as a function of this quantity for various layer separation in a system of 
$N=14$ electrons. $T^{z}=0$ corresponds to the ground state of the balanced
system at a given layer
separation, and each curve is for not too large $T^{z}$ well described by a 
parabola.
\label{fig2}}
\end{figure}

Let us now analyze the anisotropy parameter in terms of
exact-diagonalization results.
The lowest states with a given value of $T^{z}$ have vanishing total angular 
momentum on the sphere \cite{Haldane83}, i.e. they are spatially
homogeneous.
Figure \ref{fig2} shows the energy of these lowest state in the sector
of a given value of $T^{z}$ as a function of $T^{z}$ for $N=14$ electrons
and several layer separations. 
At all layer separations, in the ordered as
well as in the disordered phase, the dependence of the energy on $T^{z}$
is, for not too large $T^{z}$, parabolic, 
validating the phenomenological {\em ansatz} (\ref{anisoerg}).

\begin{figure}
\centerline{\includegraphics[width=8cm]{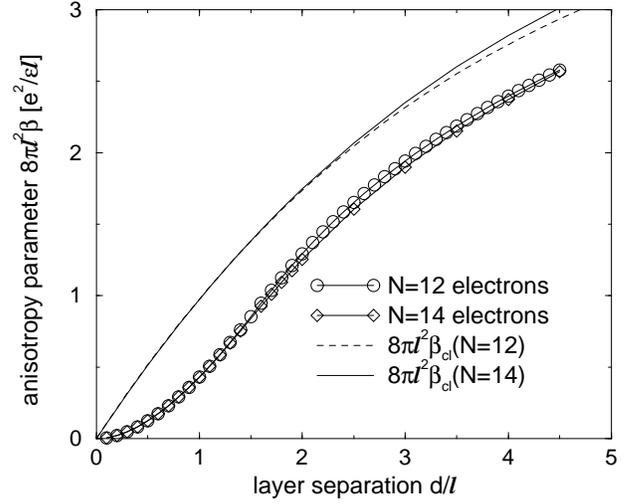}}
\caption{The anisotropy parameter $8\pi\ell^{2}\beta$ obtained from
exact-diagonalization data
as a function
of the layer separation for $N=12$ and $N=14$ electrons. Both data sets
for this (in the bulk limit) intensive quantity agree very well and show
an inflection point near the phase transition at $d/\ell\approx 1.3$. 
The corresponding
values for $8\pi\ell^{2}\beta_{cl}$ (cf. Eq.~(\protect{\ref{classbeta}}))
are also shown which describe (up to a constant) the expected behavior in the
absence of interlayer correlations.
\label{fig3}}
\end{figure}

Figure \ref{fig3}
shows values for $8\pi\ell^{2}\beta$ obtained from parabolic fits
of $\varepsilon_{a}(T^{z})$ using $T^{z}\in\{0,1,2,3\}$ 
for $N=12$ and $N=14$ electrons as a function of $d/\ell$. If higher values
of $T^{z}$ are included the quality of the fits considerably decreases. 
We therefore concentrate on
the system sizes $N=12$ and $N=14$ where a sufficient number of moderate 
values for $T^{z}$ (as compared to its maximum $N/2$) are available.
We have also plotted in figure \ref{fig3} the the classical electrostatic
expression (\ref{classbeta}) for both systems sizes.

The anisotropy parameter $\beta$ is in the bulk limit an intensive quantity.
Both data sets for $8\pi\ell^{2}\beta$
shown in figure \ref{fig3} are nearly identical establishing that $\beta$
is only very weakly dependent on the system size for already quite small
systems which are accessible via exact diagonalization techniques.
As to be expected $\beta$ increases with increasing layer
separation. Moreover it shows an inflection point near the critical value
$d/\ell=1.3$ which we interpret as a further signature of the 
quantum phase transition.
Above the inflection point the anisotropy parameter $\beta$ as obtained
from exact-diagonalization data has the same curvature as $\beta_{cl}$.
Below the inflection point at $d/\ell\approx 1.3$ both data sets differ
clearly, in particular in curvature, which indicates the presence of 
strong interlayer correlations in this regime. However, we should stress that
the concrete form of these deviations from the classical behavior, namely
the occurrence of an inflection point and a change in curvature,
is the result of the present numerical study and has not been predicted
on other theoretical grounds.

The results of subsection \ref{gsunbiased} have established the absence
of interlayer correlations in an unbiased system above the critical $d/\ell$.
If interlayer correlations vanish also in a biased system 
(with finite $T^{z}$) the anisotropy parameter $\beta$ found by exact
numerical diagonalizations should be the same as $\beta_{cl}$ up to
a rigid shift (being independent of the layer separation) arising from 
intralayer effects. As seen in figure \ref{fig3} this is for 
$d\gtrsim 1.3\ell$ to a quite good degree of approximation, but not perfectly,
the case. In particular, $\beta_{cl}$ increases with increasing system size,
while the exact-diagonalization values appear to decrease. 
The small discrepancy between $\beta$ and $\beta_{cl}$ 
(after subtracting a rigid shift) might therefore be
seen as an indication for the presence of interlayer correlations
in biased systems even above the critical layer separation of the
balanced system, as predicted recently in Ref.~\cite{Joglekar02}.
However, if so, this effect appears to be rather small, at least in the
limit of vanishing tunneling and not too large biasing, consistent with the
predictions of Refs.~\cite{Hanna97,Joglekar02}.

The value for $\beta$ at $d/\ell=1$ is by a factor of about two larger than
the effective anisotropy parameter found recently from exact diagonalization
studies of a vertical pair of parabolically 
confined quantum dots in the quantum Hall
regime \cite{EYang01}. In the latter case this effective anisotropy
parameter agrees quite reasonably with results from numerical Hartree-Fock
calculations. On the other hand, the values for $\beta$ shown in figure
\ref{fig3} agree very reasonably within a discrepancy of less than ten
percent with data reported in Ref.~\cite{Moon95} for an infinite system.
Those values were obtained from an approximate effective field theory
neglecting correlation effects beyond Hartree-Fock exchange.
Therefore the data of Ref.~\cite{Moon95}
does not show an inflection point signaling a ground state phase
transition. 

\subsection{Quantum fluctuations of the pseudospin magnetization}

In Ref.~\cite{Schliemann01} the quantum phase transition between a
pseudospin-polarized phase-coherent state and a disordered ground state
was analyzed by studying the pseudospin magnetization $\langle T^{x}\rangle$
along with their fluctuation $(\Delta T^{x})^{2}=\langle (T^{x})^{2}\rangle
-\langle T^{x}\rangle^{2}$ as a function of the tunneling gap. Here we report
on results for  $\Delta T^{x}$ at zero tunneling as a function of
$d/\ell$ in the ground state within various sectors of $T^{z}$. These
states are the absolute ground state of the system at an appropriate
bias voltage between the layers.

\begin{figure}
\centerline{\includegraphics[width=8cm]{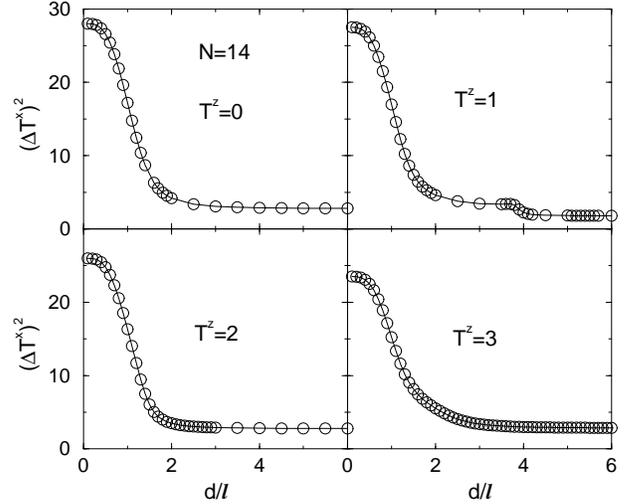}}
\caption{The pseudospin fluctuation $(\Delta T^{x})^{2}$ as a function
of the layer separation for different sectors of $T^{z}$ in a system of
$N=14$ electrons.
\label{fig4}}
\end{figure}

The ordered phase at small layer separations is characterized by large
fluctuations of the pseudospin magnetization and therefore by a large
susceptibility of this quantity with respect to interlayer tunneling
\cite{Schliemann01}.
At zero tunneling $T^{z}$ is a good quantum number while 
$\langle T^{x}\rangle=\langle T^{y}\rangle=0$, and for the
fluctuations it holds $\Delta T^{x}=\Delta T^{y}$ with $\Delta T^{z}=0$.
Figure \ref{fig4} shows $(\Delta T^{x})^{2}$ within the ground state
of several sectors of $T^{z}$ as a function of $d/\ell$ for $N=14$
electrons. At zero layer separation one has
\begin{equation}
(\Delta T^{x})^{2}=\frac{1}{2}\left(\frac{N}{2}\left(\frac{N}{2}+1\right)
-\left(T^{z}\right)^{2}\right)
\end{equation}
and for finite layer separation $(\Delta T^{x})^{2}$ decreases for all values
of $T^{z}$ with increasing $d/\ell$ to rather small values. This decay
mainly occurs in the vicinity of the critical
value $d\approx 1.3\ell$. In the upper right panel ($T^{z}=1$) yet another 
transition occurs at larger layer separations which appears to be
a peculiarity of this system size. Note that the quantity $\Delta T^{x}$
is on the other hand bounded from below by the standard uncertainty relation
$\Delta T^{x}\Delta T^{y}=(\Delta T^{x})^{2}\geq T^{z}/2$.

As seen in figure~\ref{fig4} the phase transition seems to occur rather at
the same region of $d/\ell$ in all sectors of $T^{z}$, with apparently 
a slight tendency to move to larger layer separations with increasing 
$T^{z}$. Therefore, in the case of vanishing tunneling gap,
the critical layer separation depends only very weakly on a bias voltage
between the layers. Thus, if there is an increase of the critical layer
separation in biased systems as predicted in Refs.~\cite{Hanna97,Joglekar02}, 
this effect is rather small. This is consistent with the results of the
previous subsection, and with Refs.~\cite{Hanna97,Joglekar02}.

\begin{figure}
\centerline{\includegraphics[width=8cm]{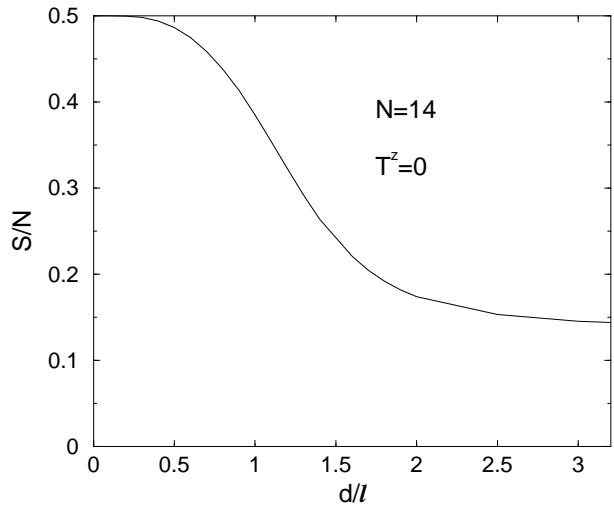}}
\caption{The effective pseudospin length $S$ per particle as
a function of $d/\ell$. This data obtained in the spherical
geometry agrees very well with recent results for a toroidal system
\protect{\cite{Nomura02}}.
\label{fig5}}
\end{figure}

Recently, Nomura and Yoshioka \cite{Nomura02} have introduced a 
parameter $S$ defined by $\langle\vec T^{2}\rangle=S(S+1)$ 
to describe the ``effective length'' of the pseudospin in a given state.
Figure~\ref{fig5} shows $S$ divided by the number of particles 
for $N=14$ electrons and $T^{z}=0$ (corresponding to the upper left
panel of figure~\ref{fig4}). This plot can be compared directly with 
data of Ref.~\cite{Nomura02} obtained in the toroidal geometry,
establishing a very good agreement between exact-diagonalization results 
on the sphere and on the torus.


\section{Conclusions}
\label{conclusions}

We have investigated ground state properties of bilayer quantum Hall systems
at total filling factor $\nu=1$ and vanishing single-particle tunneling gap by
means of exact numerical diagonalizations in finite systems.
Specifically, the ground state energy, the pseudospin anisotropy parameter,
and the quantum fluctuations of the pseudospin magnetization are studied
as  functions of the layer separation in units of the magnetic length.

The exact ground state energies are compared with results of 
finite-size Hartree-Fock calculations described in section \ref{HF}.
The availability of closed expressions for pair distribution functions and
Hartree-Fock energies even in finite systems is a specific property of 
the spherical system geometry used here. The exact ground state energies
(with a contribution from a neutralizing background being subtracted)
is independent of $d/\ell$ above the critical layer separation. This
demonstrates the decoupling of layers and the loss of spontaneous phase
coherence between them in the disordered phase. 

We have also performed a very detailed analysis of the 
effective pseudospin anisotropy parameter. 
We have found accurate numerical values
for this quantity as a function of the layer separation, and compared it
with a classical electrostatic expression valid in the absence of
interlayer correlations. This comparison establishes the strong
interlayer correlations in the ordered phase at small layer
separations, and the quantum phase transition is signaled by an
inflection point of the anisotropy parameter at the phase boundary.
Moreover, we have analyzed the possibility of 
interlayer correlations in biased
systems even above the phase boundary of the unbiased case. Certain features
of our data are not inconsistent with the occurrence of this effect, which,
however, appears to be quite small at least in the limit of vanishing tunneling
amplitude.

In summary our results show that the quantum phase transition in quantum Hall
bilayers at total filling factor $\nu=1$ shows its signatures in various 
physical quantities and represents a subtle correlation effect.

\acknowledgements{I thank Anton A. Burkov, Steven M. Girvin,
Yogesh N. Joglekar, Allan H. MacDonald, and, in
particular, Charles B. Hanna for many useful discussions.}


%

\end{document}